\documentclass[12pt]{emulateapj}

\usepackage{txfonts}

\def\ltsima{$\; \buildrel < \over \sim \;$}
\def\simlt{\lower.5ex\hbox{\ltsima}}
\def\gtsima{$\; \buildrel > \over \sim \;$}
\def\simgt{\lower.5ex\hbox{\gtsima}}
\def\farcm{\hbox{$\mkern-4mu^\prime$}}

\def\farcs{\hbox{$^{\prime\prime}$}~}

\def \feh {[Fe/H]}

\newcommand{\bb}{\hbox{\it B\/}}
\newcommand{\vv}{\hbox{\it V\/}}

\newcommand{\bmv}{\hbox{\bb--\vv\/}}

\shorttitle{On the absolute age of M71}
\shortauthors{Di Cecco et al.}

\begin{document}
\title{On the absolute age of the metal-rich globular M71 (NGC~6838): I. optical photometry}

\author{   
A. \,Di Cecco\altaffilmark{1,2}, 
G. \, Bono\altaffilmark{3,4}, 
P.G.\, Prada Moroni\altaffilmark{5,6}, 
E. \, Tognelli\altaffilmark{3,6},      
F. \, Allard\altaffilmark{7}, 
P.B. \, Stetson\altaffilmark{8} 
R. \, Buonanno\altaffilmark{1,3}, 
I. \, Ferraro\altaffilmark{4}, 
G. \, Iannicola\altaffilmark{4}, 
M. \, Monelli\altaffilmark{9,10}, 
M. \, Nonino\altaffilmark{11}, 
L. \, Pulone\altaffilmark{4}
}

\altaffiltext{1}{INAF - Osservatorio Astronomico di Teramo, via M. Maggini, I-64100, Teramo, Italy}
\altaffiltext{2}{INAF - ASI Science Data Center, via del Politecnico, I-00133, Rome, Italy}
\altaffiltext{3}{Universit\'a di Roma Tor Vergata, via della Ricerca Scientifica,  I-00133, Rome, Italy}
\altaffiltext{4}{INAF - Osservatorio Astronomico di Roma, Via  di Frascati 33, I-00040 Monte Porzio Catone, Rome, Italy}   
\altaffiltext{5}{Universit\'a di Pisa, Largo Pontecorvo 3, I-56127, Pisa, Italy}
\altaffiltext{6}{INFN-Pisa, Largo Pontecorvo 3, I-56127, Pisa, Italy}
\altaffiltext{7}{CNRS - Centre de Recherche Astrophysique de Lyon, UMR 5574, Université de Lyon, École Normale Supérieure de Lyon, 46 Allée d'Italie, F-69364 Lyon Cedex 07, France}
\altaffiltext{8}{Dominion Astrophysical Observatory, Herzberg Institute of Astrophysics, National Research Council, 5071 West Saanich Road, Victoria, BC V9E 2E7, Canada}
\altaffiltext{9}{Instituto de Astrof\'isica de Canarias, Calle Via Lactea, E-38200 La Laguna, Tenerife, Spain}
\altaffiltext{10}{Departamento de Astrof\'isica, Universidad de La Laguna,
38200 La Laguna, Tenerife, Spain}
\altaffiltext{11}{INAF - Osservatorio Astronomico di Trieste, via G.B. Tiepolo 11, I-40131, Trieste, Italy}

\date{\centering drafted \today\ / Received / Accepted }

\begin{abstract}
We investigated the absolute age of the Galactic globular cluster M71
(NGC~6838) by using optical ground-based images ($u',g',r',i',z'$) collected
with the MegaCam camera at the Canada-France-Hawaii-Telescope (CFHT).
We performed a robust selection of field and cluster stars by applying a new
method based on the 3D ($r',u'-g',g'-r'$)  Color-Color-Magnitude-Diagram.
The comparison between the Color-Magnitude-Diagram of the candidate cluster stars and a new set of isochrones, at the locus of the Main Sequence Turn Off (MSTO), suggests an absolute age of 12$\pm$2~Gyr.
The absolute age was also estimated using the difference in magnitude
between the MSTO and the so-called main sequence knee,
a well defined bending occurring in the lower main sequence.
This feature was originally detected in the near-infrared (NIR) bands
and explained as a consequence of an opacity mechanism (collisionally induced
absorption of molecular hydrogen) in the atmosphere of cool
low-mass stars \citep{bono2010}. The same feature was also detected
in the $r'$, $u'-g'$ and in the $r',g'-r'$ CMD, thus supporting previous
theoretical predictions by \citet{bory97}. The key advantage
in using the $\Delta^\mathrm{Knee}_\mathrm{TO}$ as an age diagnostic is that it is
independent of uncertainties affecting the distance, the reddening and
the photometric zero-point. We found an absolute age of 12$\pm$1~Gyr
that agrees, within the errors, with similar age estimates, but the
uncertainty is on average a factor of two smaller. We also found
that the $\Delta^\mathrm{Knee}_\mathrm{TO}$ is more sensitive to the metallicity
than the MSTO, but the dependence becomes vanishing using the
difference in color between the MSK and the MSTO.

\end{abstract}

\maketitle

\section{Introduction}
\label{Intro}

The Galactic Globular Cluster (GGCs) M71 --NGC~6838-- \citep[\feh =-0.78~dex,][]{harris96} is a very interesting stellar 
system  since it belongs to the small sample of metal--rich GCs present 
in our Galaxy \citep{harris96} and in the Local Group \citep{cezario13}. 
Moreover, it is located relatively close, and indeed its distance is smaller 
than 4~Kpc \citep{grund02}.    
This means that M71 is a fundamental laboratory to constrain the evolution 
of old, metal-rich, low--mass stars \citep{hodder92}. The absolute age 
of M71 can also play a key role in the occurrence of a bifurcation in 
the age-metallicity relation of inner and outer halo GGCs recently 
suggested by \citet{dotter11}.   
Moreover, M71 together with a few more metal--rich GGCs  
(47 Tuc, NGC~6528, NGC~6553) plays a fundamental role in the calibration 
of spectrophotometric indices adopted to constrain the age, the metallicity 
and the shape of the Initial-Mass-Function of unresolved stellar populations 
in early type galaxies 
\citep{boselli09, cappellari12, conroy12, conroy13, spiniello12}.  
 
The main drawback of M71 is that it is affected by a large reddening with a 
mean value\footnote{The mean value has been calculated by the online tool 
available at http://irsa.ipac.caltech.edu/applications/DUST/.} of 
$<E(\bmv)>$=0.323$\pm$0.017~mag \citep{schlegel98} and by a differential 
reddening with a mean value of $<\delta~E(\bmv)>$=0.035$\pm$0.015~mag, which 
extents up to the maximum value of  $\delta~E(\bmv)$=0.074~mag 
\citep{bonatto13} \citep[for the reddening values see also][]{kron76,frogel79,hodder92,kraf03}. This is 
mainly due to the fact that M71 is located at very low Galactic latitude 
(l=56$^{\circ}$.75, b=-4$^{\circ}$.56).    

The literature concerning the age estimates of M71 is quite rich. 
By comparing optical CCD photometry with the oxygen enhanced models 
of \citet{bergbush91}, \citet{hodder92} found that M71 has an age 
in good agreement with the 14$\pm$2 and 16$\pm$~2~Gyr isochrones. 
They also found that M71 is coeval with 47~Tuc, while by using the 
same theoretical models \citet{geffert2000} suggested for M71 an 
older age (18~Gyr).  By calibrating the cluster distance with the Hipparcos field
subdwarfs parallaxes, and by adopting a reddening of $E(\bmv)$=0.28~mag, \citet{reid98}  
found a true distance modulus of $\mu$=13.19$\pm$0.15~mag 
and an absolute age of 8$\pm$1~Gyr (as estimated from his 
Fig.~10a). He also suggested that 47~Tuc, within the uncertainties, is coeval 
(10$\pm$1~Gyr).
 
The Hipparcos catalog was also used with Stro\"emgren photometry by \citet{grund02}, who 
derived a true distance modulus of 
$\mu$=12.84$\pm$0.04$\pm$0.1\footnote{We used the \citet{cardelli89} 
law to unredden the original value of $M_V$=13.71~mag.}~mag 
by adopting $E(\bmv)$=0.28~mag. By comparing the M71 data with 
the isochrones from \citet{vandenberg00}, the authors also provided an 
absolute age of 12~Gyr and found the same age for 47~Tuc.

After the transformation of the \citet{clem2} M71 fiducial sequence to the SDSS bands with
the equations provided by \citet{tucker06}, and assuming $E(\bmv)$=0.28~mag, \citet{an09} 
found a distance modulus of $\mu$=12.86$\pm$0.08~mag and $\mu$=12.96$\pm$0.08~mag respectively
according to the metallicity scale provided by \citet{kraf03} and \citet{cg97}.

To further improve the temperature sensitivity around the MSTO, 
\citet{brasseur10} adopted optical-NIR colors and, by using a true distance modulus of 
$\mu$=13.16~mag and a mean cluster reddening of $E(\bmv)$=0.20~mag, they found an 
absolute age of 11~Gyr. This age estimate was based on a new set of cluster isochrones 
computed by \citet{vandenberg12}.
More recently, \citet[][hereinafter V13]{vandenberg13} by using homogeneous optical 
photometry based on WFC/ACS images, provided accurate relative ages for 55 Galactic 
globulars. They adopted an improved vertical method and a new approach to fit 
Horizontal Branch (HB) stars with predicted Zero Age Horizontal Branch models. 
They found for M71 an absolute age of 11.00$\pm$0.38~Gyr.

In a more recent paper, by investigating the age of 61 GCs, 
\citet[][hereinafter LVM13]{leaman13} found that  for abundances \feh$\geq$1.8~dex, 
the GCs obey to two different Age--Metallicity--Relations (AMRs). In particular, 
they found that one--third of the entire sample is, at fixed cluster age, 
systematically more metal--rich by 0.6~dex. 
Moreover, they suggest that the bulk of the 
metal--rich sequence formed in situ in the Galactic disk, while a significant 
fraction of the metal--poor globulars formed in dwarf galaxies that were 
subsequently accreted by the Milky Way.   
However, the current scenario concerning the absolute age distribution of GCs is far from being settled. 
In a similar investigation \citet{marin09}, by using homogeneous photometry for 
a sample of 64 globulars, found that the bulk of GCs are coeval, 
the spread in age being smaller than 5\%, and do not display an AMR. They also found 
a small group of systematically younger clusters that display an AMR similar to 
the globulars associated to the Sagittarius dwarf galaxy. To take account of the 
observed age distribution, they suggested that the old globulars formed in situ, 
while the young ones formed in dwarf galaxies that were later accreted.    

V13 and LVM13 did not include in their sample reddened bulge GCs, but on the basis of the AMR relation they found for metal--rich 
clusters they predicted steadily decreasing ages for more metal--rich bulge clusters 
(see their section 7.3). In particular, for the prototype metal--rich bulge cluster 
NGC~6528 they predicted an age of 10~Gyr. On the other hand, \citet{lagioia}
by using WFC/ACS and UVIS/WFC3 images that cover a time interval of 10 years 
performed an accurate proper motion separation between field and cluster stars in 
NGC~6528. On the basis of optical photometry they found that NGC~6528 is old (12~Gyr) 
and coeval with 47~Tuc. 
        
The above evidence indicates that we still lack solid constraints on the age distribution 
of metal--rich GCs. In this investigation we address the absolute age 
of M71 by adopting optical ground--based (SDSS bands) collected with MegaCam at Canadian--France--Hawaii--Telescope (CFHT). The layout of the paper is the following.       \textsl{•}
In section \ref{obs} we present the dataset and the strategy adopted for data reduction. 
In  section \ref{sect2} we discuss the approach we adopted to identify candidate cluster 
and field stars. The theoretical framework adopted to estimate the absolute age of 
M71 is presented in section \ref{sect4} and \ref{sect5}, while in section \ref{sect6} we discuss the absolute age 
estimates based on the classical main sequence turn-off and on the main sequence knee. 
Finally, in section \ref{sect7} we summarize the results and outline future perspectives.

\section{Observations and data reduction} \label{obs}

\begin{figure*}[t]
\includegraphics[width=14cm]{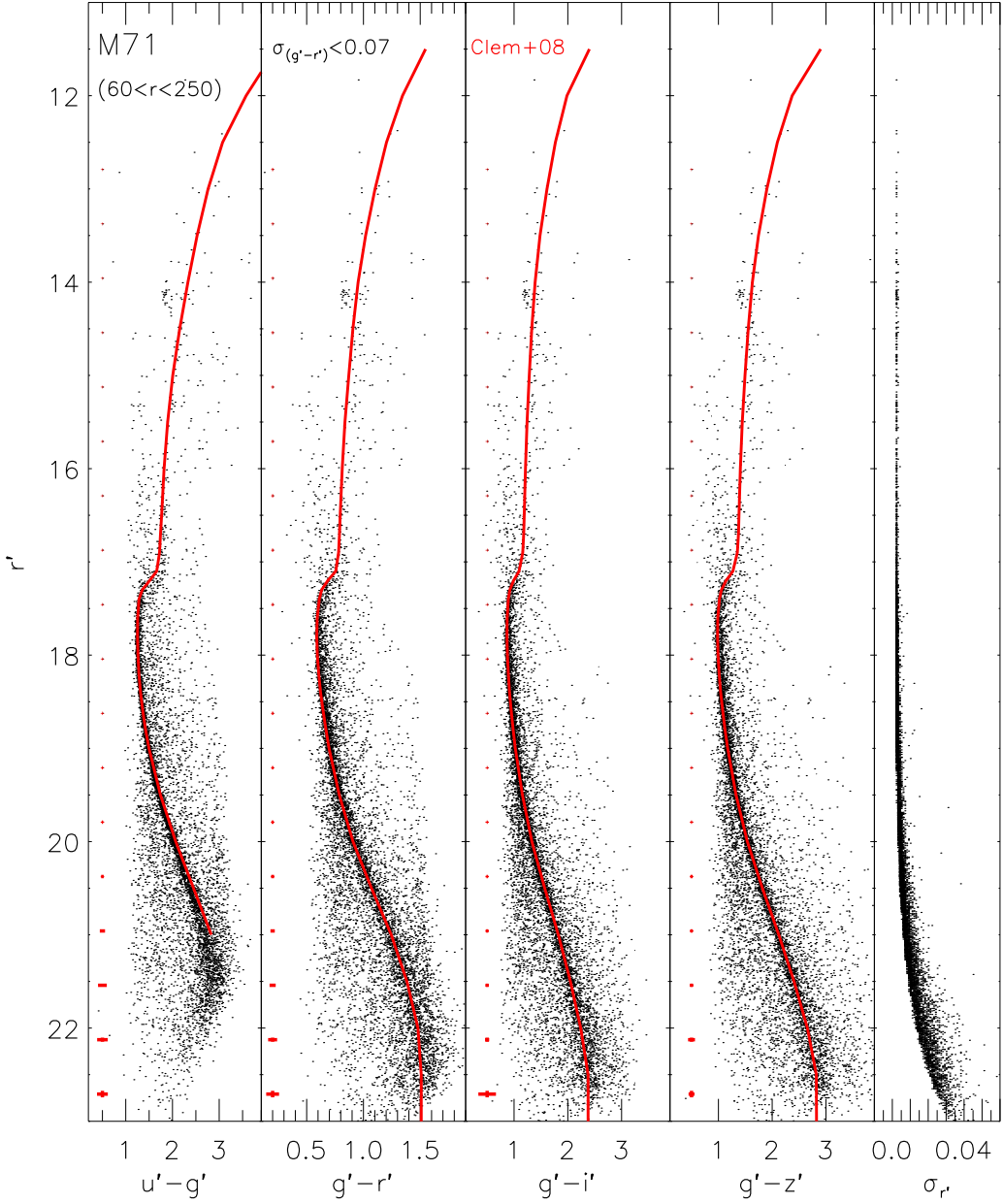}
\caption{From left to right: four CMDs of M71 based on optical images collected with the MegaCam at CFHT. Data plotted in the different CMDs were selected using 
photometric quality parameters ($\chi\approx1$, $|sharp|<1$, $\sigma_{g'-r'}<0.07$) and according to the radial distance 60\farcs$<r<250$\farcs. 
The red lines display the fiducial lines provided by \citet{clem2}, while the red bars plotted on the left side take account of the photometric error both in magnitude and in color. The sample of stars located at $r'$$\sim$14.2, $u'-g'$$\sim$1.75~mag are cluster red HB stars. Data plotted in the rightmost panel display the photometric error in the $r'$-band.}
\label{f1}
\end{figure*}

To investigate the absolute age of M71 we used the images of the MegaCam 
(36 Charge Coupled Devices, CCD; total field of view 
(FoV):$1^{\circ}$x$1^{\circ}$; scale: 0.187$\farcs$/pixel) mounted 
on the CFHT.
The dataset includes 50 dithered images\footnote{The current 
optical images were acquired during two nights: July 8 and 13, 2004; 
proposal IDs: 04AC03, 03AC16; P.I. J. Clem; observing program: 
'CFHT Star Cluster Survey'.} taken with the $u'g'r'i'z'$ (SDSS) bands 
\citep{fukugita}. We retrieved from the Canadian Astronomy Data Centre,   
five shallow and five deep images per filter with the following exposure 
times (ETs): 
$ET(u')$=30, 500~sec; $ET(g',r')$=5, 250~sec; $ET(i')$=5, 300~sec; and 
$ET(z')$=15, 500~sec. The mean seeing ranges from $\sim$1.0 ($u',g',z'$) 
up to $\sim$1.3 ($i',r'$) arcseconds. The data were pre-processed with 
the ELIXIR programs \citep{magnier}.
The globular M71 covers a small sky area, and indeed its tidal radius is 
$r_t$=8~\farcm.9 \citep{harris96}, therefore, we only used the 20 
innermost CCDs for a total FoV of 0$^{\circ}$.6x1$^{\circ}$.0.
For each frame we provided an accurate PSF (Point Spread Function) photometry 
by using the DAOPHOT~IV and ALLSTAR \citep{stetson1}, and we used 
DAOMATCH/DAOMASTER to scale individual chips on a common geometrical reference 
system. Once we obtained the global catalog (master list of stars) of the 
entire dataset we run ALLFRAME \citep{stetson2} simultaneously over the 
entire set of images. We obtained a list of 370,000 stars with at least one 
measurement in four different photometric bands. 
To calibrate the instrumental magnitudes, we used the local standard stars 
($\sim$6,000) provided by \citet{clem1}. To validate the adopted 
transformations, Fig.\ref{f1} shows the comparison in four different 
Color-Magnitude-Diagrams (CMDs) between the current photometry and the 
ridgelines (red solid lines) provided by \citet{clem2}. They agree quite 
well not only along the Red Giant Branch (RGB), but also along the 
Main Sequence (MS). 
 
The anonymous referee noted that for $r'$-band magnitudes fainter than 
$\sim$22~ mag ($r',g'-r'$ CMD; $r'\sim$20.5 mag for the $r',u'-g'$ CMD) the 
ridgelines provided by Clem attain colors that are slightly bluer (redder for 
the $r',u'-g'$ CMD) than observed. The difference is mainly caused by the 
different approaches adopted to perform the photometry, in selecting candidate 
cluster stars and in the calibration to standards. Our approach is based on 
simultaneous photometry on shallow and deep $u',g',r',i',z'$ images. 
This allowed us to reach a better photometric precision 
in the faint magnitude limit. Stars plotted in Fig.~1 have been selected 
according to the cluster radial distance (60$\le$$r$$\le$250 arcsec), and to 
photometric quality parameters ($\sigma_{g'-r'}=0.07$~mag, 
$\chi\approx1$, $|sharp|<1$). 
Data plotted in this figure indicate that the lower MS is better defined, less 
affected by the field star contamination (see Fig.~12 in Clem et al. 2008).          
We adopted the same local standard stars provided by \citet{clem1} and 
followed a similar procedure \citep{clem2} to transform the instrumental 
magnitudes into standard magnitudes. Unfortunately, we cannot perform a 
detailed comparison among the different calibration equations, since their  
zero-points and coefficients of the color terms are not available. 

The intrinsic error in magnitude and in color are 
plotted as red error bars. They take account of the photometric error and 
of the absolute calibration and attain values of the order of a few 
hundredths of magnitude down to the lower main sequence. This trend 
is also supported by the photometric error in the $r'$-band plotted 
in the rightmost panel of the same figure. It is of the order of a few 
hundredths of magnitude down to the bending of the MS ($r' \sim$23 mag).  

Finally, we performed an astrometric solution of the catalog by using the 
UCAC4 catalog. The root mean square error of the positions is 
$\sigma \sim$55 $10^{-3}$ arcseconds.  

\section{Identification of field and cluster stars} \label{sect2}

\begin{figure*}[t]
\centering
\includegraphics[width=12cm,height=11cm]{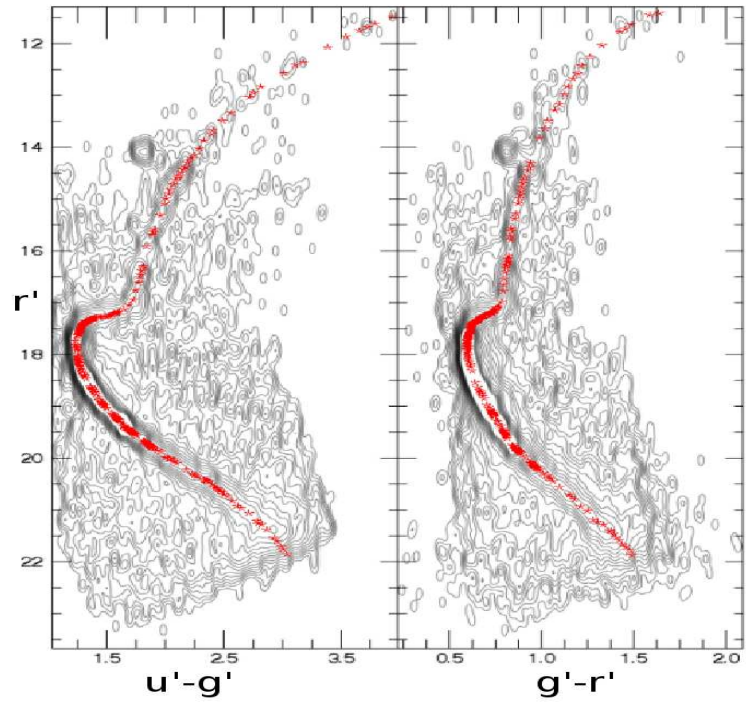} 
\caption{Left -- Iso-density contours (black lines) of the entire data set in 
the $r'$,$u'-g'$ CMD. The ridgeline of the candidate cluster stars is overplotted 
with red asterisks.
Right -- same as the left, but for the $r'$,$g'-r'$ CMD. 
}
\label{f2}
\end{figure*}

To separate candidate cluster and field stars we devised a new approach that 
fully exploit the multi-band data set we are dealing with. 
We performed a preliminary radial selection and to avoid the crowded central 
regions we only selected stars located inside an annulus with $r$=2~\farcm 
and $r$=5~\farcm centred on the cluster. On the basis of this sample we 
generated two isodensity maps in the $r',u'-g'$ and in the $r',g'-r'$ 
CMDs (see Fig.~\ref{f2}). To properly constrain the cluster 
ridge line we developed a numerical algorithm that pin points the peaks of 
the isocontour plots and provides a preliminary version of the cluster 
ridgeline. Note that in the approach we devised, the overdensity caused by 
red HB stars was neglected. The preliminary ridgeline is then fit with a 
bicubic spline and visually smoothed, in particular in the bright portion 
of the RGB. The analytical ridgelines are sampled at the same 
$r'$ magnitude levels (see Table~\ref{tab1}). The red lines plotted in the left and in the right 
panel of Fig.~\ref{f2} show the final version of the estimated ridgelines.

\begin{figure*}[t]
\centering
\includegraphics[width=13cm,height=11cm]{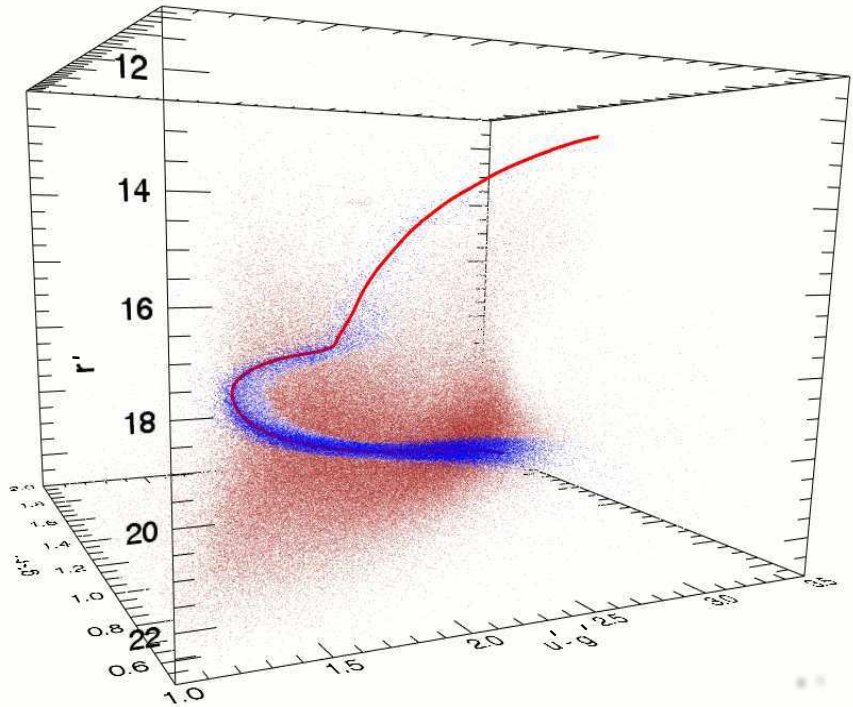} 
\caption{$r'$,$u'-g'$,$g'-r'$ Color-Color-Magnitude-Diagram (CCMD) of M71. 
The blue dots are the candidate cluster stars selected along the 3D ridgeline 
(red line). The red dots are candidate field stars.}
\label{f3}
\end{figure*}

The two analytical ridgelines were used to generate a 3D plot 
Color-Color-Magnitude Diagram (CCMD)-- $r',u'-g',g'-r'$ --of the cluster. 
The Fig.~\ref{f3} shows the 3D ridgeline (red solid line) together with 
the entire sample of stars with measurement in the the $u',g',r'$ bands. 
We considered as candidate cluster stars those located within the 
3D ridgeline $\pm$1$\sigma$, where $\sigma$ is defined as two times 
the quadratic sum of the photometric errors in the three quoted bands.

The above approach has several indisputable advantages when compared with 
classical photometric methods in selecting field and cluster stars.  

{\em i}) The selection in the CCMD takes advantage of the typical 
effective temperature correlation of stellar structures in a 
color--color plane. Current approach takes also advantage of the fact 
that we are simultaneously using photometric bands covering a broad 
range in central wavelenghts, and in particular, of the $u'$--band.  
This means a solid separation not only between stars and field galaxies, 
but also between field and cluster stars \citet{bonocarina}. 
However, the color-color-diagram approach is prone to degeneracy between dwarf 
field and giant cluster stars sharing very similar optical colors.   

{\em ii}) The selection in the CMD takes advantage of the typical 
correlation between cluster stars located at the same distance 
(apparent magnitude) and their color. However, the CMD approach 
is prone to spurious selections interlopers located at different 
distances, i.e. field stars associated either to the Galactic halo 
or to the Galactic disk stars. This means that after the selection 
of the stars located across the ridgeline we need to take account of the 
fraction of field stars that have been erroneously classified as 
candidate cluster stars \citep[see][]{dicecco13}  
  
\begin{figure*}
\centering
\includegraphics[width=14cm, height=13cm]{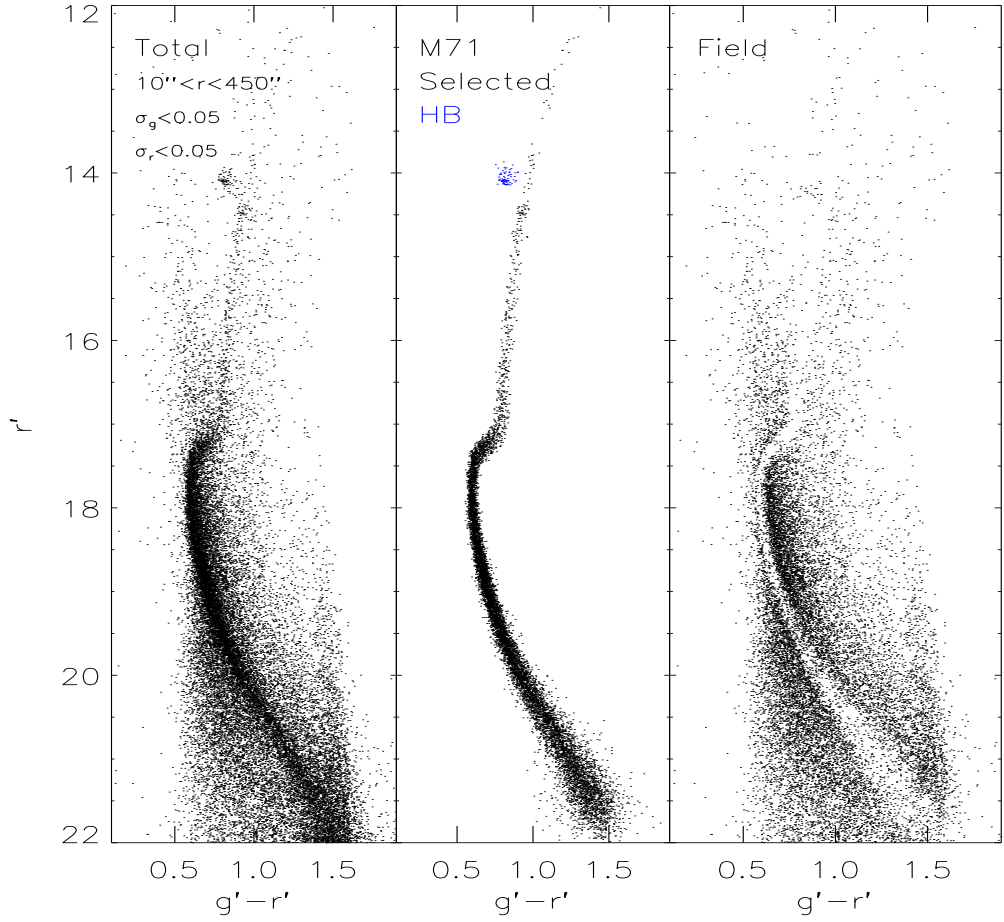}
\caption{Left -- $r'$,$g'-r'$ CMD of the entire photometric catalog. Stars 
plotted in this CMD were selected in photometric error ($\sigma_g<0.05, 
\sigma_r<0.05$) and in cluster radial distance (10\farcs $<r<$ 450\farcs). 
Middle -- Same as the left, but for candidate cluster stars. The blue dots 
display cluster red HB stars.  
Right -- Same as the left, but for candidate field stars. 
}
\label{f4}
\end{figure*}

The approach based on the CCMD takes account of the advantages of 
both the color-color-diagram and of the CMD selection criteria. To validate the approach 
we adopted, Fig.~\ref{f4} shows the $r',g'-r'$ CMD, before (left) and after (middle) 
the selection. The candidate field stars are plotted in the right panel. 
The magnitude and color distribution of field stars support the plausibility 
of the adopted selection criterion, since we recover the two typical peaks 
in $g'-r'$ color at $\sim$0.6 and at $\sim$1.5~mag of field dwarf stars 
\citep{ivezic08}.   
Data plotted in the middle panel display a smooth and well 
sampled CMD from the tip of the RGB down to the lower MS. 
Moreover, the presence of field stars in the same region of the CMD 
in which we would expect candidate cluster stars on the basis of the CMD, 
is further supporting the approach we developed. 

 We highlight that the isodensity maps plotted in Fig.~2 show a secondary peak at magnitude 
slightly brighter (17$\le r' \le$17.5~mag) and bluer ($u'-g'\sim$1.2 mag; 
$g'-r' \sim$ 0.6~mag) than the ridgeline of the candidate cluster stars. 
We performed several tests changing the ridgeline either in the $r',u'-g'$ 
or in the  $r',g'-r'$ density map and checking the impact on the CCMD. We found that this blue plume mainly includes candidate field 
stars. This finding is supported by the smooth distribution of cluster 
MSTO stars in Fig.~1 and by the presence of the blue plume stars in the 
CMD of candidate field stars plotted in the left panel of Fig.~4. We thank 
the anonymous referee for drawing our attention on this group of stars. 

\section{Theoretical framework and cluster isochrones} \label{sect4}

To estimate the cluster age we used the evolutionary tracks and the cluster isochrones of the Pisa Stellar Evolution Data Base\footnote{\url{http://astro.df.unipi.it/stellar-models/}} for low-mass stars, computed using FRANEC stellar evolutionary code \citep{deglinnocenti08, tognelli11}. The input physics and the physical assumptions adopted to construct the evolutionary tracks have already been discussed in \citet{database2012}, while an analysis of the main theoretical uncertainties affecting stellar tracks and isochrones is provided in \citet{valle13a,valle13b}. Among the available grids of models, we take account of those computed with an $\alpha$-enhanced chemical mixture ([$\alpha$/Fe]=+0.3) and the recent solar heavy--element mixture by \citet[][As09]{asplund09}.
Recent spectroscopic estimates give for M71 an iron abundance of \feh=-0.82$\pm$0.02 dex
\citep[][Ca09]{carretta09}.  
This abundance is based on a reference solar iron abundance 
of $\log \epsilon_\odot(\mathrm{FeI})$ = 7.54. To transform it into the 
\citet{asplund09} solar iron abundance ($\log \epsilon_\odot(\mathrm{Fe})$=7.50), 
we used the following relationship:
\begin{eqnarray}
[\mathrm{Fe/H}]_\mathrm{As09} &=& [\mathrm{Fe/H}]_\mathrm{Ca09} + \log \epsilon_\odot(\mathrm{Fe})_\mathrm{Ca09} - \log \epsilon_\odot(\mathrm{Fe})_\mathrm{As09} = \nonumber\\ &=& [\mathrm{Fe/H}]_\mathrm{Ca09} + 0.04
\end{eqnarray}
This means an iron abundance for M71 of \feh=-0.78 dex, and by assuming 
an $\alpha$--enhancement of [$\alpha$/Fe] = +0.3 dex, a global metal abundance 
per unit mass of $Z$=0.0037 and a primordial helium abundance of $Y$=0.256. 
The closest metallicities for $\alpha$-enhanced isochrones currently available 
in the Data Base are: $Z$=0.002, $Y$=0.252; $Z$=0.003, $Y$=0.254 and $Z$=0.004, $Y$=0.256. The helium 
abundances as a function of the metal abundances were estimated by using a 
linear helium--to--metal enrichment ratio of $\Delta Y / \Delta Z$ = 2 \citep{fukukawa}. 
The above chemical compositions imply iron abundances of \feh=-1.04, \feh=-0.87, and \feh=-0.74 dex, 
respectively. The adopted sets of evolutionary models bracket, within the errors, the observed iron abundance of M71.  

The two adopted grids of evolutionary models cover the typical range in mass of 
low--mass stars, namely $M$= 0.30--1.10 $M_{\sun}$. Moreover, they were constructed 
by adopting three different values of the mixing length parameter: $ml$=1.7, 1.8, and 1.9. 
The corresponding cluster isochrones cover the age range from 8 to 15~Gyr.
The luminosities and the effective temperatures provided by the evolutionary models 
were transformed into magnitudes and color indexes by using the synthetic spectra 
by \citet{brott05} for 2000 K $\le \mathrm{T}_{eff} \le 10\,000$ K and by 
\citet{castelli03} for $10\,000\,\mathrm{K} \la \mathrm{T}_{eff} \la 50\,000$ K. 
The transformations into the observational plane follow the prescriptions discussed 
in \citet{girardi02}.

To compute the AB magnitudes in the $u'g'r'i'z'$ photometric system we used the 
USNO40 Response Functions\footnote{They are available at the following url 
http://www-star.fnal.gov/ugriz/Filters/response.html, see for more details 
\citep[see][]{smith02}} with an air mass of 1.6. This is the air mass value 
at which the observations of the local standard stars in M71 provided by 
\citet{clem1} were performed.

\begin{figure*}
\centering
\includegraphics[width=19cm, height=13cm]{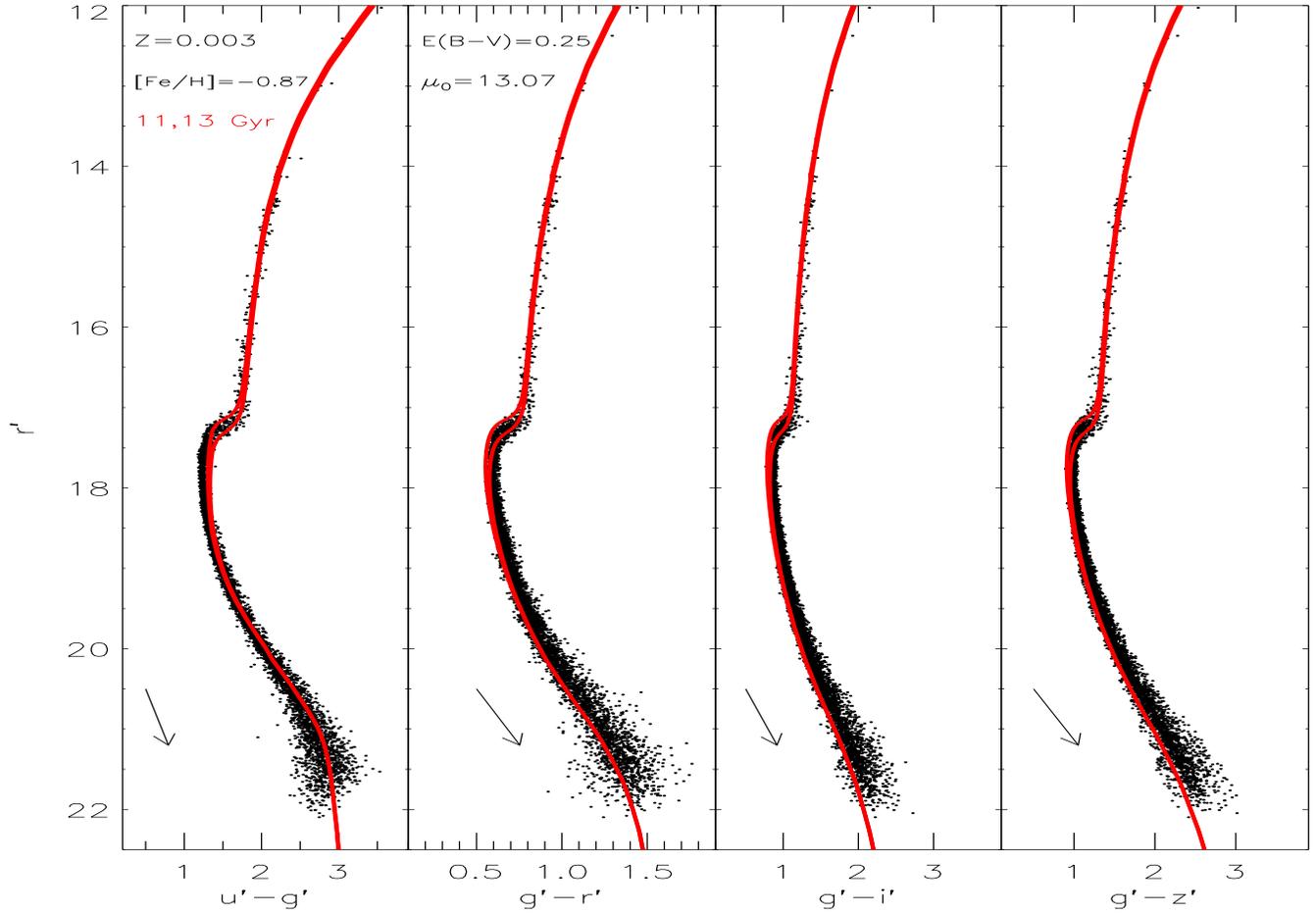}
\caption{From left to right comparison between theory and observations in 
different CMDs of M71. Black dots display candidate cluster stars selected in 
photometric error and in cluster radial distance. The red lines display two 
$\alpha$--enhanced cluster isochrones at fixed metal content (Z=0.003, Y=0.254) 
and ages of 11 and 13~Gyr, respectively. The true distance modulus and the 
cluster reddening adopted to overplot evolutionary prescriptions are also 
labelled.  The reddening in the different bands was estimated using the 
semi-empirical Cardelli's relation. The black arrows show the reddening 
vectors in the different CMDs for an arbitrary change in cluster reddening. 
}
\label{f5}
\end{figure*}

Fig.~\ref{f5} shows the comparison between candidate cluster stars and the 
set of isochrones constructed assuming by $Z$=0.003, $Y$=0.254 and a $ml$=1.9. 
The cluster isochrones were plotted by adopting a true distance modulus of 
 $\mu$=13.07~mag and a cluster reddening of $E(\bmv)$=0.25~mag. The adopted 
values agree quite well with similar estimates available in the literature 
(see Table~\ref{tab2}). The selective absorptions in the individual bands were 
estimated by assuming a ratio of absolute to selective extinction of 
$R_V$=3.1 and the empirical reddening law by \citet{cardelli89}.  
The extinction ratios were computed by adopting the effective 
wavelengths of the $u'g'r'i'z'$ photometric system  provided by 
\citep[][]{smith02}. For the five adopted bands we found the 
following selective absorption ratios: 
$A_{u'}/A_V$=1.58, $A_{g'}/A_V$=1.20, $A_{r'}/A_V$=0.87, 
$A_{i'}/A_V$=0.66, and $A_{z'}/A_V$=0.49 \citep[see also][]{diceccoage}. 
The black arrows 
plotted in the bottom left corners display the reddening vectors in the 
four different CMDs. The two red lines display the cluster isochrones 
for 11 and 13~Gyr. Theory and observations agree quite well not only 
along the evolved sequences (RGB; Sub Giant Branch, SGB), but also along the MS. There 
is a mild evidence that cluster isochrones in the lower main sequence 
($r'\le$20.5 mag) become slightly bluer than observed stars in 
the $r'$, $g'-r'$ (panel $b$), $r'$, $g'-i'$ (panel $c$) and in the 
$r'$, $g'-z'$ (panel $d$), but the agreement across the turn-off 
regions is quite good in the four different CMDs. 
Similar discrepancies between observed and predicted colors have 
also been found in the optical and in the 
NIR bands \citep{kuci05}. 
The only exception is the $r'$, $u'$-$g'$ CMD (panel $a$), 
in which the observed stars in the MSTO region ($r'\sim$17.8 mag) 
are a few hundreds of magnitude bluer than predicted. A similar 
discrepancy was also found by \citet{an09} using $\alpha$--enhanced 
cluster isochrones computed with YREC code \citep{sills, delahaye} 
and transformed into the observational plane by adopting MARCS 
stellar atmosphere models \citep{marcs}. They suggested that 
the main culprit could be missing opacities in the model 
atmospheres at short wavelengths, as well as the color transformation 
for the highly reddened cluster M71. 
Current findings support the discrepancy, but the comparison 
also display a good agreement along the MS ($r'\le$20 mag).  
Thus suggesting that the mismatch between theory and observations 
might also be due to marginal changes in the adopted chemical 
composition.  

\begin{figure*}
\centering
\includegraphics[width=19cm, height=13cm]{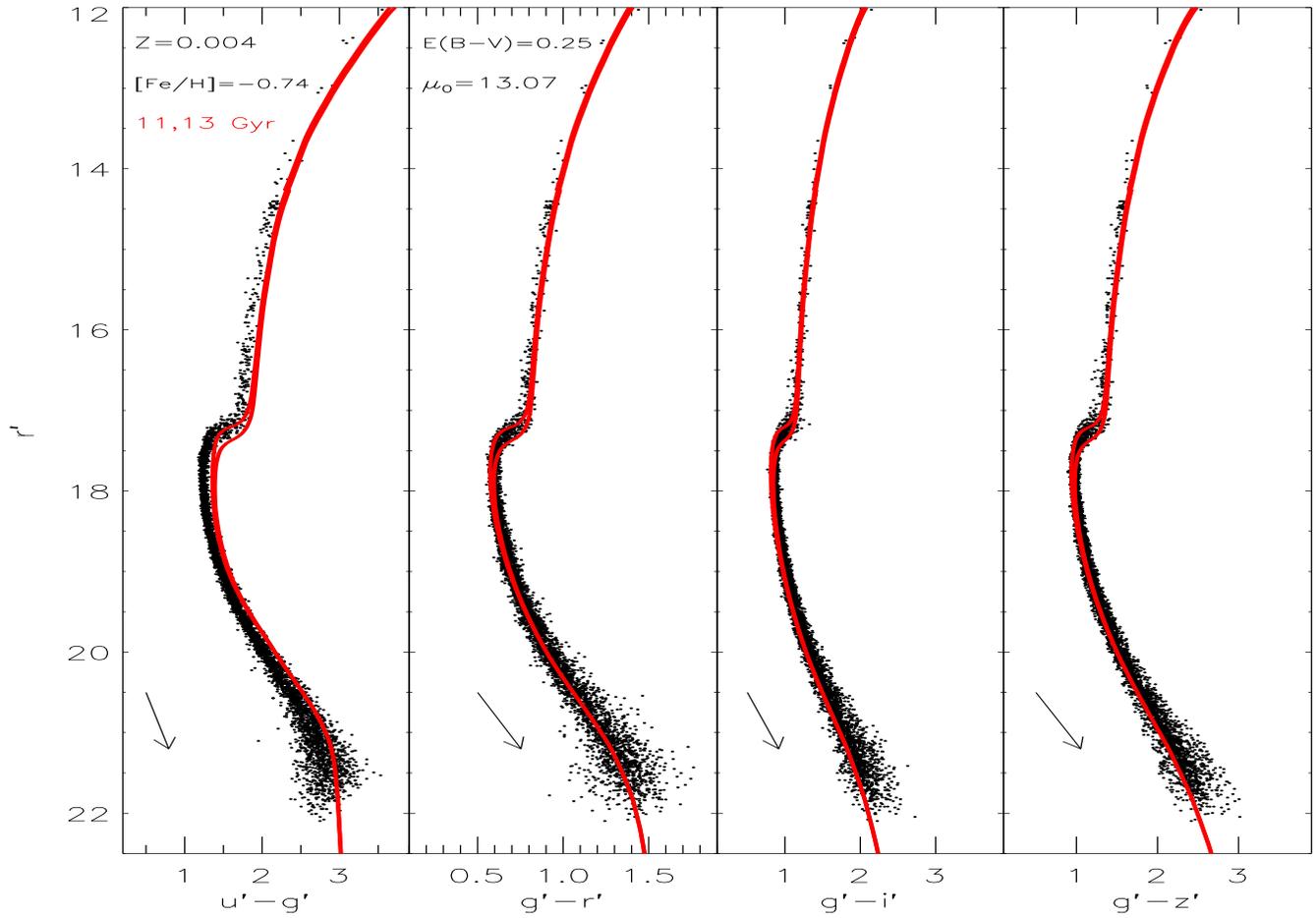}
\caption{Same as Fig.~5, but the comparison is with two more 
metal--rich (Z=0.004, Y=0.256) $\alpha$--enhanced cluster isochrones.}
\label{f6}
\end{figure*}

To further constrain the above working hypothesis, we performed 
the same comparison, but by adopting a slightly more metal--rich 
chemical composition ($Z$=0.004, $Y$=0.256). Data plotted in 
Fig.~\ref{f6} clearly show that the discrepancy between theory 
and observations increases. This applies not only to the MSTO 
region where the difference is of the order of 0.3~mag, but also
along the RGB/SGB and the MS, and indeed the isochrones attain colors 
that are systematically bluer than observed. Moreover, it is worth mentioning 
that the increase of 0.13~dex in iron abundance also causes a clear 
discrepancy along the SGB region in the other CMDs (panels $b$, $c$, $d$) 
and does not improve the agreement in the lower MS.

Current evidence further supports the strong sensitivity of the $u'$--band 
to the metal content and suggests that the mild discrepancy found in 
the color of TO region of the $r'$, $u'-g'$ CMD might also be affected 
by the adopted abundance of $\alpha$-elements and/or of CNO-elements.

\section{Constraining the nature of the knee in the lower MS} \label{sect5}

The occurrence of a well defined knee in the low--mass regime of the MS 
has already been brought forward in the literature. It has already been 
detected in several old ($\omega$~Cen, Pulone et al. 1998; 
M4, Pulone et al. 1999, Milone et al. 2014; NGC~3201, Bono et al. 2010; 
47~Tuc, Lagioia et al. 2014; NGC~2808, Milone et al. 2012) and intermediate-age 
\citep{sara09} stellar systems and in the Galactic bulge 
\citep{zoccali00} by using NIR CMDs. 
Current evolutionary prescriptions indicate that this feature is mainly 
caused by the collisionally induced absorption of H$_2$ at 
NIR wavelengths \citep{saumon94}. The shape of the bending marginally 
depends on the metal content, but the magnitude of the knee is essentially 
independent of cluster age and of metallicity. 

The above theoretical and empirical evidence suggests that the magnitude 
difference between the MS knee (MSK) and the MSTO, i.e. 
$\Delta^\mathrm{Knee}_\mathrm{TO}$, is a robust diagnostic to constrain 
the absolute cluster age.
The key advantages of the new approach is that the above difference in 
magnitude is independent of uncertainties affecting both the distance 
and the reddening of the stellar system. Recent empirical evidence 
indicate that the this new approach provides cluster ages that are 
a factor of two more precise when compared with the classical method 
of the MSTO \citep{bono2010, sara09}. 

In passing, we 
also note that the $\Delta^\mathrm{Knee}_\mathrm{TO}$ method appears more robust than the 
traditional horizontal (difference in color between the RGB, typically 
2.5~mag brighter than the MSTO, and the MSTO ) and vertical 
(difference in magnitude between the HB luminosity level, 
typically at the RR Lyrae instability strip, and the MSTO) methods 
(\citet{stetson99,buonanno98, marin09}; V13). According to theory, 
MS stellar structures with a stellar mass of $M\sim$0.5-0.4~$M_\odot$ are minimally affected by 
uncertainties in the treatment of convection, since the convective 
motions are nearly adiabatic \citep{saumon08}. The same outcome 
does not apply to cool HB stars and to SGB stars adopted 
in the vertical and in the horizontal method, respectively. Moreover, 
the MSK can be easily identified in all stellar systems with a well 
populated MS. It is independent of the uncertainties affecting 
the estimate of the HB luminosity level when moving from metal--poor 
(blue and extreme HB morphology) to metal--rich (red HB morphology) 
GCs (\citet{cala07, iannicola09}; V13). 
Morover, the $\Delta^\mathrm{Knee}_\mathrm{TO}$ method is also independent of the 
theoretical uncertainties plaguing the color--temperature transformations 
required in the horizontal method to constrain the cluster age. 

In this context it is worth mentioning that both the vertical and the 
horizontal method are robust diagnostics of optical CMDs. Recent 
accurate and deep NIR CMDs show that HB stars display a well defined 
slope when moving from hot and extreme HB stars to cool red HB stars 
\citep{delp06, coppo11, milone13, stetson14}. The same outcome applies to the CMDs based 
on near UV and far UV bands \citep{ferraro12}. This means that 
the identification of the HB luminosity level required by the vertical 
method is hampered by the exact location in color of the anchor along 
the HB. Moreover, the difference in color between the MSTO and an 
anchor along the RGB, required by the horizontal method, is hampered 
by the fact that the RGB becomes almost vertical in NIR bands. This 
means that the difference in color is steadily decreasing 
\citep{coppo11, stetson14} 

Finally, the minimal 
dependence on the metal content allow us to tight correlate the   
$\Delta^\mathrm{Knee}_\mathrm{TO}$ directly to the absolute age of the stellar 
system. This means that the new approach provides absolute age 
estimates, while the horizontal and the vertical method do provide 
estimates of the relative age.

\subsection{The impact of CIA on the knee of the lower MS} \label{sect5.1}

To further constrain, on an empirical basis, the robustness of the  
$\Delta^\mathrm{Knee}_\mathrm{TO}$ to estimate the absolute age of GCs we also 
investigated the beding in the optical bands \citep{bono2010}
and in particular in the $u'$-- and in the $g'$--band. The reason 
is twofold. 
1) Detailed atmosphere models taking account of the collision-induced 
absorption (CIA) opacities of both H$_2$--H$_2$ and H$_2$--He indicate 
that CIA induces a strong continuum depression in the NIR region of 
the spectrum \citep{bory97,bory00,bory02}. The same outcome applies 
to the UV region of the spectrum for surface gravities and 
effective temperatures typical of cool MS stars \citep[see Figs. 5 and 6 
in][]{bory97}. It is interesting to note that the above 
effect is also anticorrelated with the iron abundance. 
The increase in the iron abundance causes, at low effective temperatures, 
an increase of the molecular opacities \citep[see Fig.~7 in][]{bory97}, 
leading to a weaker impact of the CIA on the continuum.  
2) Interestingly enough, recent deep and accurate optical--UV CMD 
already showed the presence of a well defined knee in the low--mass 
regime of the MS. \citet{an08} clearly identified a well defined 
knee in the lower MS of two old open cluster: NGC~2420 and M67 (see 
their Fig.~16). A similar evidence was also found in the V,B$-$I CMD 
of M4 \citep{stetson14}. The lack of a clear evidence of a knee 
in the lower MS can be associated either to the use of V,R,I bands 
in which the knee is less evident or to the fact that the photometry 
is not deep/accurate enough to properly identify the knee.  

To properly constrain the impact that CIA has on the UV flux, and in 
particular on the blue magnitude and colors, we performed a specific test. 
We adopted an old (12~Gyr) cluster isochrone and selected three stellar 
structures with stellar masses ranging from 0.30 to 0.45 $M_\odot$. The 
surface gravities and the effective temperature of the selected models 
are listed in Table~\ref{tab3}. 
Using these values as input parameters, we computed two specific sets 
of atmospheric models (spectra), one that neglects and one that takes 
account of the CIA opacities. 

\begin{figure*}
\centering
\includegraphics[width=12cm, height=9cm]{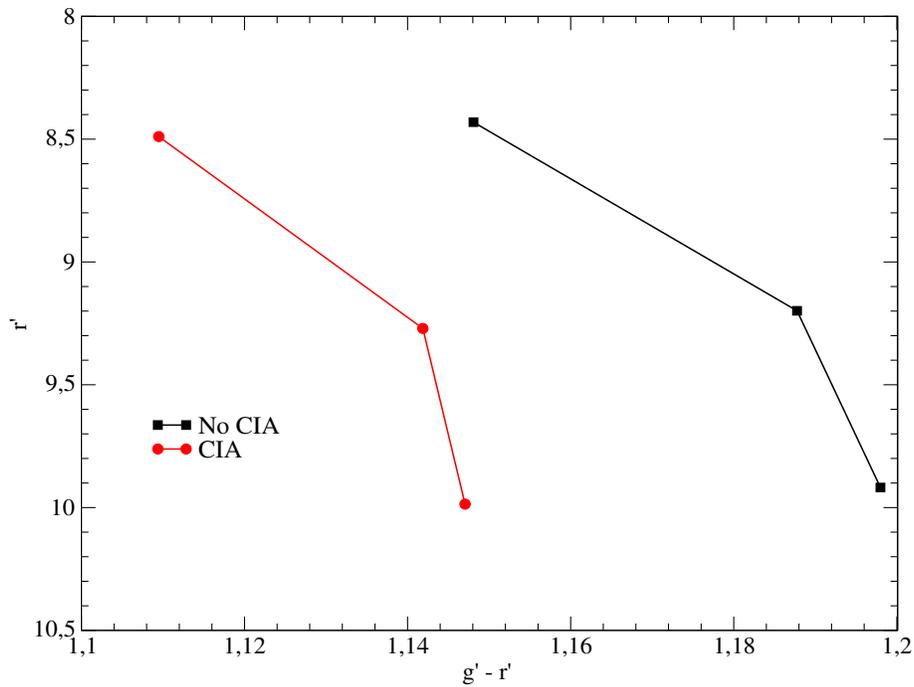}
\caption{Comparison between cluster isochrone (12~Gyr) transformed into the observational  
plane adopting stellar atmosphere models either including (red line) or 
neglecting (black line) CIA opacity. 
}
\label{f7}
\end{figure*}

Data plotted in Fig.~\ref{f7} display that CIA has a substantial impact on 
the emerging flux of high-gravity, low effective temperature stellar 
structure. 

 In the color-magnitude region common to the two sets, the models that take 
account of CIA (see red line in Fig.~\ref{f7}) are, at fixed stellar mass, 
$\sim$0.1 mag fainter than those without CIA opacity, the differences 
increasing as the effective temperature decreases.
Notice also that, at a fixed luminosity, the inclusion of CIA produces
models $\sim$0.05~mag bluer than the models without CIA. The difference 
becomes more evident in the very low-mass regime and at longer wavelengths. 

In this context it is worth mentioning that the above atmosphere models either 
taking account of or neglecting the CIA opacity rely on the input physics 
adopted in the latest version of the PHOENIX BT-Settl atmosphere structures 
\citep{allard11}. These models when compared with previous atmosphere 
models computed by \citet{brott05} present two differences: 
a) the former use the solar mixture provided by Asplund et al. (2009), 
while the latter use the solar mixture provided by Grevesse \& Noels (1993);     
b) the former models include a more complete list of CIA opacities. 
In particular, they take account of molecular hydrogen (H$_2$), molecular 
nitrogen (N$_2$), methane and carbon dioxide (CO$_2$). The main difference with 
previous models is in the use of opacities provided by Abel \& Frommhold (2011) 
that provide less absorptions at high temperatures when compared with earlier 
computations by \citet{fubory}. 
A more detailed list of the adopted CIA opacities used in the current 
BT-Settl atmosphere models is given in Table~\ref{tab4} together with their 
references.   

Note that the current isochrones have been transformed into 
the observational plane using the homogeneous set of atmosphere models 
provided by \citet[][see Sect. \ref{sect4} for more details]{brott05}. To 
constrain the difference between the magnitudes and colors based on the 
\cite{brott05} and \cite{allard11} sets of atmosphere models we performed a test. 
We transformed the best fit cluster isochrone (12~Gyr) and chemical composition  
(\feh=-0.87 dex, [$\alpha$/Fe]=+0.3) using also the recent \citet{allard11} 
spectra\footnote{\citet{allard11} spectra are available at the following 
URL: \url{http://phoenix.ens-lyon.fr/Grids/BT-Settl/}.}. 

To overcome problems in the adopted solar mixtures the two sets of 
atmosphere models were interpolated at fixed metallicity (\feh=-0.87 dex). 
We found that the difference in the $r'$ magnitude is negligible, 
and indeed, it is of the order of 1\% at MSTO and becomes at most 
of the order of a few hundredths of magnitude in the region across 
the MSK. We performed the same experiment, but the atmosphere models 
were interpolated at fixed global metallicity Z. 
The difference between the two sets of atmosphere models was, once again, 
minimal. Thus suggesting that new atmosphere models have a minimal impact 
on the conclusions of the current investigation.     
A more detailed investigation of the the impact that CIA and molecular 
line opacities have on UV, optical and NIR colors over a broad range of 
metal abundances will be addressed in a forthcoming paper.      

%
%
%

\section{The absolute age of M71} \label{sect6}

Current photometry is very accurate and precise over at least ten magnitudes 
(see Fig.~5). This outcome applies in particular to the bluer bands 
($u',g',r'$). Moreover, theory and observations display a well defined knee 
($r'\sim$21.5 mag) in the $r'$, $u'-g'$ ($u'-g'\sim$2.9 mag) and in 
the $r'$, $g'-r'$ ($g'-r'\sim$1.4 mag) CMD. 
The knee was also detected in the $r',g'-r'$ CMD by 
\citet[][see their Fig.~12]{clem2} using the same data set, but a different 
data reduction strategy. The plausibility of the detection was also supported 
by the knee showed by  metal--rich  cluster isochrones (VandenBerg et al. 2006) 
adopted by \citet[][see their Fig.~6]{clem2} in the $r',u'-g'$ CMD for the 
typical absolute age ($\sim$12~Gyr) and metal--abundance (\feh$\sim$-0.71 dex) of M71.  
A similar evidence, but based on multiband (SDSS) observations, was brought forward 
for metal--rich clusters by \citet[][see their Figures 5 and 8]{an09}.

Therefore, we decided to take advantage of this evidence and to estimate 
the absolute age of M71 by using both the classical MSTO method and the 
new $\Delta^\mathrm{Knee}_\mathrm{TO}$ method. The latter method appears very promising, 
since the vertical method might be more prone to possible systematic 
uncertainties in the metal--rich regime. 
 The HB morphology of metal--rich GCs is typically characterized by a stub of
red stars. These stars in several optical and optical-NIR CMDs appear tilted 
\citep[][V13]{raimondo02, milone12b, lagioia}.  
Moreover, metal--rich GCs--alike M71--harbour either no RR Lyrae or at most 
a few (only one in 47~Tuc). If present they are quite often either evolved 
\citep{bonorr2003} or peculiar \citep{pritzl2000}. 
The vertical method is typically anchored to the mean magnitude of cluster 
RR Lyrae. Therefore, the application to metal-rich clusters 
is--once again--difficult, since a theoretical/empirical correction is 
required. The homogeneity of the diagnostic adopted to estimate the absolute 
age is a crucial issue in this context, since we are interested in constraining  
possible differences in age between metal--poor and metal--rich GCs. 

\begin{figure*}
\centering
\includegraphics[width=19cm, height=13cm]{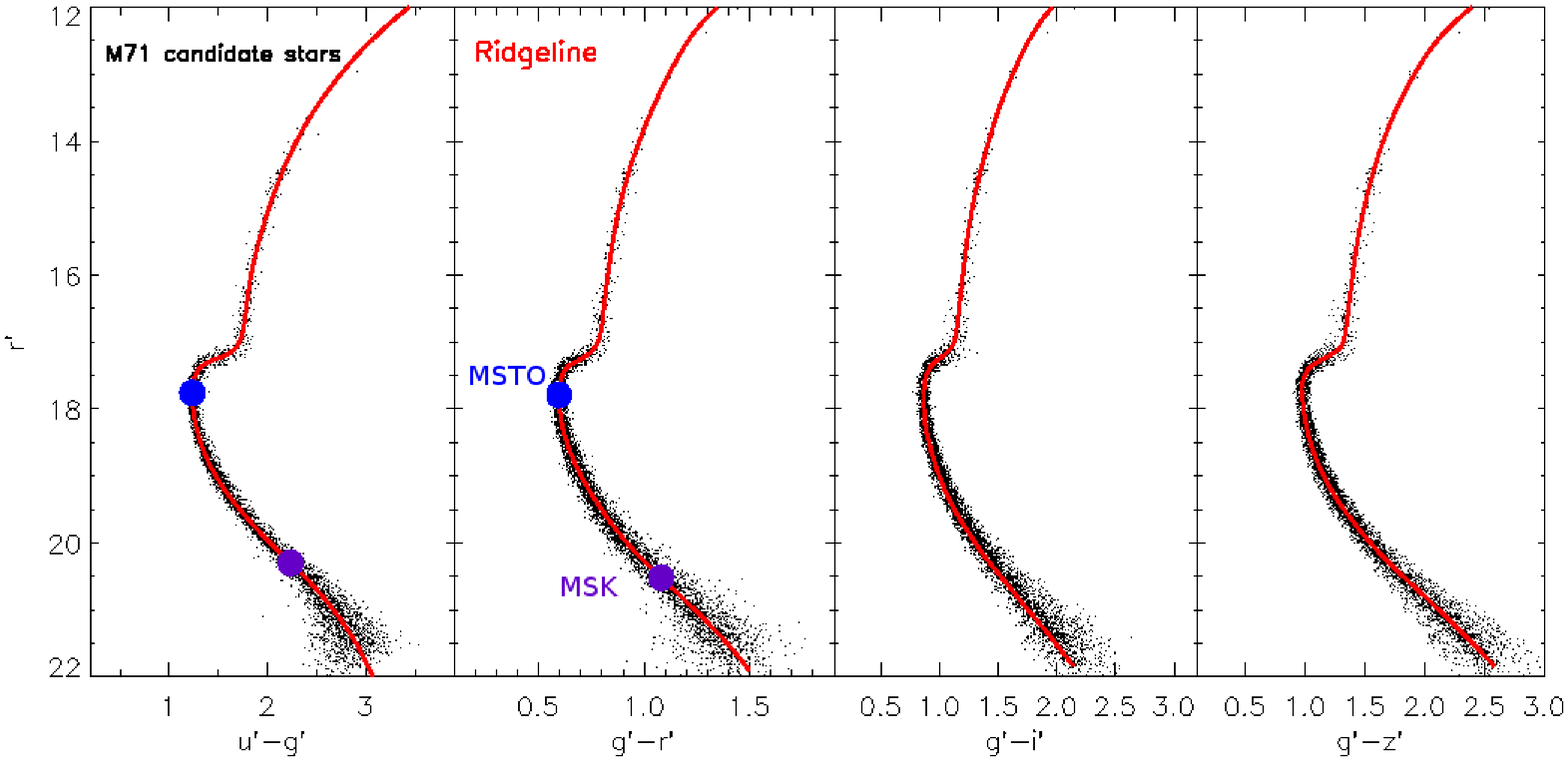}
\caption{From left to right optical CMDs of candidate cluster stars. Stars 
plotted in the different CMDs were selected according to photometric error 
and cluster radial distance. The red lines display the ridgelines, while the 
two large blue filled circles plotted in the $r',g'-r'$ CMD mark the position 
of both the MSTO and of the MSK.
}   
\label{f8}
\end{figure*}

The approach adopted to estimate the observed magnitudes of both MSTO and MSK 
has already been discussed in detail by \citet{bono2010} in their investigation 
of NGC~3201. In the following, we briefly outline the main steps of the quoted
approach. The observed ridgeline of the $r',g'-r'$ CMD was equally sampled 
($\Delta_{r'}=$0.01 mag) using a cubic spline, and then, the MSTO and the MSK magnitudes were estimated as the points showing the minimum curvature along the 
MS ridgeline.  We found that the MSTO is located at 
$r'=17.79\pm0.01$~mag, while the MSK at $r'=20.66\pm0.01$~mag. 
Note that the uncertainties account for the photometric errors and for 
errors in the ridgeline and in the location of the above points 
(see Fig.~\ref{f8}). 

To estimate the absolute age of M71 with the classical MSTO method we adopted 
the more metal--poor ($Z$=0.003, $Y$=0.254) set of isochrones, since they take 
account for the observed stars in the different CMDs. Note that the marginal 
difference in color between observations and theory--in the 
$r'$, $u'-g'$ CMD--has a minimal impact on the absolute age estimate, 
since we are only using the magnitude of the MSTO. 

\begin{figure*}
\centering
\includegraphics[width=14cm,height=12cm]{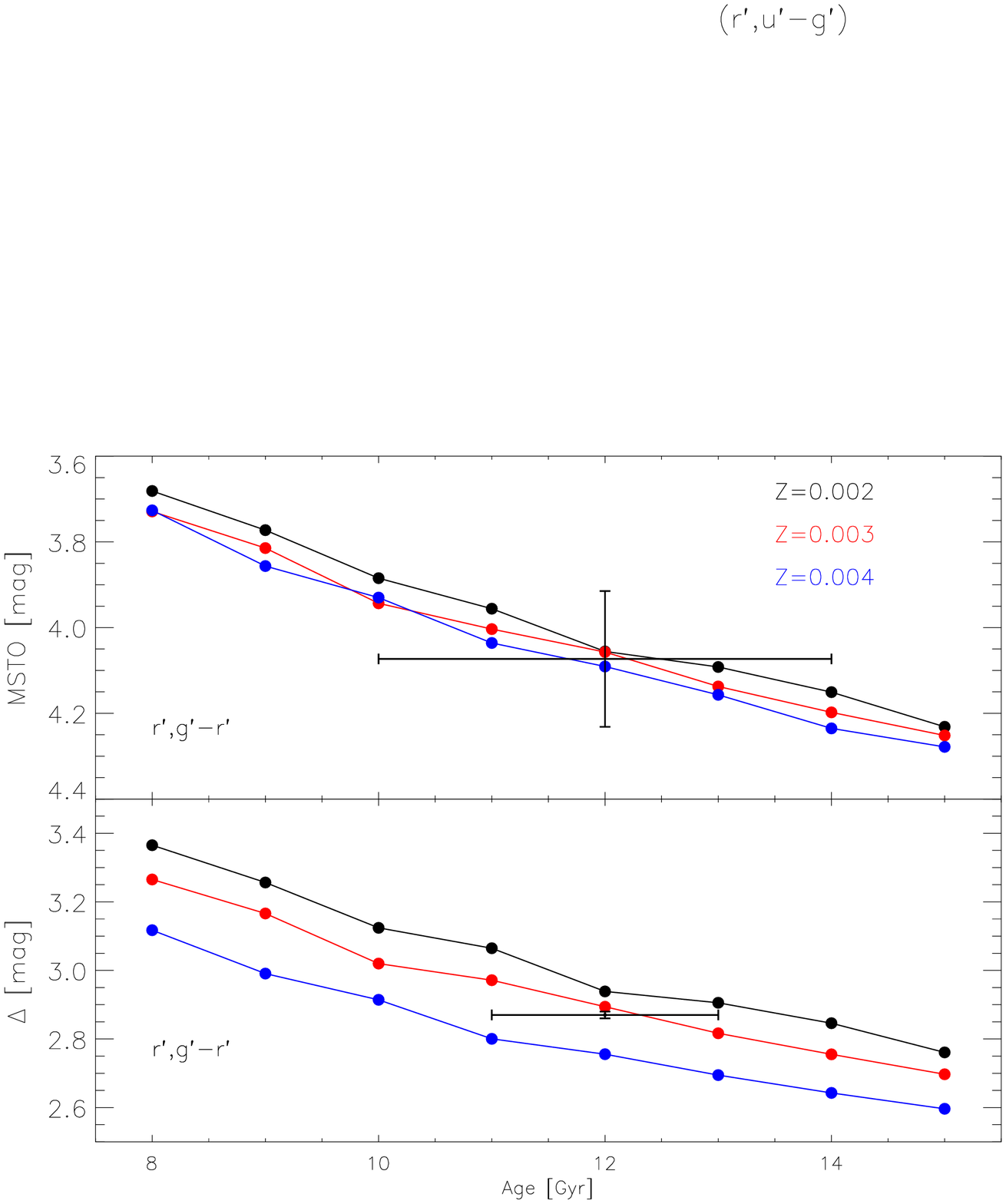}
\caption{Top -- Predicted absolute magnitude of the MSTO as a function of cluster age. 
The MSTO was selected in $r'$,$g'-r'$ isochrones constructed by adopting 
three different $\alpha$--enhanced chemical compositions (see labeled values). 
The observed unreddened MSTO value is 4.07~mag and it is shown (by the vertical bar) considering the uncertainties of the photometry and of the distance modulus and reddening. The horizontal bar shows the estimated uncertainty on the age. 
Bottom -- Same as the top, but for the $\Delta^\mathrm{Knee}_\mathrm{TO}$ parameter (2.87 mag). The vertical bar only takes account of the photometric error. The horizontal bar shows the estimated uncertainty on the age. 
} 
\label{f9}
\end{figure*}

Data plotted in our Fig.~\ref{f5} indicate an absolute age for M71 of order of 12~Gyr. 
To properly estimate the age and the errors affecting the ages based on the MSTO 
we followed the approach suggested by \citet{renzini88} \citep[see also][]{buonanno98}. 
We selected the $r'$-band absolute magnitude of the MSTO on cluster isochrones 
for ages ranging from 8 to 14~Gyr. These estimates were performed in three sets 
of $r'$,$g'-r'$ isochrones computed assuming different chemical compositions, 
namely: $Z$=0.002, $Y$=0.252; $Z$=0.003, $Y$=0.254 and $Z$=0.004, $Y$=0.256. Note that 
we adopted this CMD, because these are the bands in which both theory and 
observations are more precise.  On the basis of the quoted estimates we 
performed a linear regression among age, metallicity and $r'$-band absolute 
magnitude of MSTO. We found the following relation: 

\begin{eqnarray}
\log t[\mathrm{Gyr}] &=& (-1.06\pm0.06) + (0.51\pm0.01)\times M^{TO}_{r'} - \nonumber\\
&&(0.09\pm0.01)\times [\mathrm{Fe/H}] 
\end{eqnarray}

where $t$ is the absolute age in Gyr and the other symbols have their usual 
meaning. 
Using the above relation we estimated an absolute age for M71 of 
12$\mp$2~Gyr ( see top panel of Fig.\ref{f9} ).   
The error budget of this determination takes account of photometric errors 
in the magnitude of the MSTO ($\sigma_{r'} \sim$0.01-0.02~mag), in the cluster 
reddening ($\sigma_{E(\bmv)} \sim$0.05~mag), in the true distance modulus 
($\sigma_\mu \sim$ 0.15~mag) and in the chemical composition 
($\sigma_{[M/H]} \sim$0.1~dex). The current estimate is, within the errors, 
in good agreement with the absolute ages for M71 available in the literature 
(see Table~\ref{tab2}). The agreement is quite good with recent estimates based 
both on ground-based \citep{grund02, brasseur10} and 
on space (V13) photometry together with up-to-dated 
cluster isochrones \citep{vandenberg12}.

To estimate the absolute age with the method of the $\Delta^\mathrm{Knee}_\mathrm{TO}$ we adopted the same sets of 
cluster isochrones ($Z$=0.003) we used for the MSTO method. However, the cluster 
isochrones with ages ranging from 8 to 14~Gyr were equally sampled using 
a cubic spline.  The same algorithm adopted to estimate the observed MSTO and the MSK 
on the ridgelines was also adopted on the splined isochrones. The observed and predicted 
differences in magnitude ($\Delta^\mathrm{Knee}_\mathrm{TO}$) between the MSK and the 
MSTO ($r', r'-g'$) are plotted in the bottom panel of Fig.~\ref{f9}. The predicted 
$\Delta^\mathrm{Knee}_\mathrm{TO}$, at fixed metal content, plotted in this figure 
show a well defined linear trend over the investigated age range. 
To provide solid constraints on the errors affecting the absolute age 
based on the $\Delta^\mathrm{Knee}_\mathrm{TO}$, we followed the same approach adopted 
for the MSTO. Using the $r'$,$g'-r'$ cluster isochrones, we obtained the 
following relation:
\begin{eqnarray}
\log t[\mathrm{Gyr}] &=& (2.18\pm0.06) - (0.48\pm0.02)\times \Delta^\mathrm{Knee}_\mathrm{TO} - \nonumber\\
&&(0.30\pm0.02)\times [\mathrm{Fe/H}] 
\end{eqnarray}
where the symbols have their usual meaning. 
Using the above analytical relation, we found an absolute age for M71 of
12~Gyr. The error on the current age is $\pm$1~Gyr and takes account of
uncertainties in photometry and in iron abundance. The above error is
roughly a factor of two smaller than the typical error of absolute ages
based on the MSTO. The difference is due to the fact that the $\Delta^\mathrm{Knee}_\mathrm{TO}$
diagnostic is independent of uncertainties affecting both the distance modulus
and the cluster reddening. Moreover, the difference in
magnitude--$\Delta^\mathrm{Knee}_\mathrm{TO}$--is also minimally affected by uncertainties
in the absolute photometric zero-points. This means that this approach can
also provide very accurate estimates of the relative ages of stellar
systems.

In this context, it is worth mentioning that the coefficient of the age
indicator attains, as expected, very similar values between MSTO and
$\Delta^\mathrm{Knee}_\mathrm{TO}$. This is the consequence of the fact that the above
diagnostics rely on the same age indicator: the MSTO.
On the other hand, the dependence on the iron abundance is larger for the
$\Delta^\mathrm{Knee}_\mathrm{TO}$ than for the MSTO (0.3 vs 0.1 dex). The difference is
independent of the adopted magnitude ($r'$) and color ($g'-r'$).
We performed several tests using the same cluster isochrones and we found
that the metallicity dependence is strongly correlated to the definition
of the MSK (maximum vs minimum curvature). Moreover, we also found that
the metallicity dependence of the age diagnostic ($\Delta^\mathrm{Knee}_\mathrm{TO}$)
becomes similar to the MSTO using the the difference in color instead
of the difference in magnitude. The above evidence suggests that the
precision of the different age diagnostics is a complex balance among
photometric and spectroscopic precision age sensitivity together with
uncertainties affecting distances moduli and reddening corrections.
We plan to provide a detailed mapping of the sensitivity of optical,
optical-NIR and NIR CMDs in a forthcoming investigation.

\section{Summary and final remarks} \label{sect7}

We performed accurate and deep multiband ($u',g',r',i',z'$) photometry of the 
metal--rich Galactic globular M71 (NGC~6838). The images were collected with 
MegaCam at CFHT and cover one square degree around the center of the cluster.  
The image quality and the strategy adopted to perform the photometry on both 
shallow and deep exposures allowed us to provide precise photometry in all 
the bands from the tip of the RGB to five magnitudes fainter than the MSTO. 
The extended data set of local standards provided by \citet{clem1} 
allowed us to fix the absolute zero--point of the above bands with a 
precision that is, on average, better than  0.02 mag.  

The selected cluster is projected onto the Galactic bulge. This means 
that absolute age estimates are hampered by the contamination of field 
bulge stars. To overcome this thorny problem we deviced a new approach 
based on the ($r'$,$u'-g'$,$g'-r'$) Color-Color-Magnitude Diagram. First
we derived accurate ridgelines of candidate cluster stars in the 
$r'$,$u'-g'$ and in the $r'$,$g'-r'$ CMDs using iso-density contours. 
Then, they were combined to provide a 3D trace to properly selected 
candidate cluster stars. The key advantage of this approach is that 
it takes advantage of both intrinsic properties (color selections) 
and of the clustering (same distance). The results appear very 
promising, but the approach can be further improved using optical-NIR 
color selections. 

We computed specific sets of cluster isochrones covering a broad 
range in ages and in the adopted chemical composition. Moreover, 
to properly fit observed red giant branch stars, the evolutionary 
models were computed by adopting three different values for the 
mixing length parameter. To constrain possible systematics in the 
transformation into the observational plane, 
we paid special attention on the adopted stellar atmosphere models 
and in the adopted-pass bands. We found that the latter plays a crucial 
role in the comparison between theory and observations.   

We also investigated the impact that the collisional induced absorption 
(CIA) opacity has in occurrence of the main sequence knee in the Sloan 
bands. We found that CIA affects not only NIR magnitudes, but also 
blue bands such as the $u'$ and the $g'$ band. The sensitivity of the 
short wavelength regime on the CIA opacity was predicted by 
\citet{bory97} and now soundly confirmed by both theory 
and observations.      

The comparison between theory and observation is quite good over 
the entire magnitude range covered by observations in the different 
CMDs. There is evidence of a mild 
difference--a few hundredths of magnitude--between cluster isochrones 
and observations. However, it is not clear whether the difference is 
due either to the adopted atmosphere models or to the adopted 
reddening law.    

We estimated the absolute age of M71 using two different 
diagnostics. Using the MSTO and the cluster isochrones in the 
$r',g'-r'$ CMD, we found an age of 12$\pm$2~Gyr. The error takes 
account of uncertainties in cluster distance, reddening and iron 
abundances together with photometric uncertainties. 
We performed the same estimate using the knee-method \citep{bono2010}, 
i.e. the difference in magnitude between the MSTO and the MSK, in 
the same CMD. We found a cluster age that is identical to the age 
based on the MSTO, but the uncertainty is of a factor of two smaller. 
The difference is mainly due to the fact that the adopted 
diagnostic---$\Delta^\mathrm{Knee}_\mathrm{TO}$---is independent of uncertainties 
in cluster distance and reddening. However, it is more prone to  
uncertainties affecting cluster iron abundance.

The above age estimates support, within the errors, recent age 
estimates of metal--rich Galactic globulars. One of the most 
metal--rich Galactic globular--NGC~6528--was investigated 
by \citet{lagioia} using deep ACS and WFC3 images. 
Using the MSTO and a similar set of cluster isochrones they 
found and absolute age of 11$\pm$1~Gyr. The absolute age of 
this cluster was independently confirmed by 
\citet{cala14} using for the same cluster accurate 
ground--based Stro\"emegren and NIR photometry (t=11$\pm$1~Gyr) .  
The outcome applies to the age estimates of metal-rich 
bulge clusters provided by \citet{zocca03,zocca04}, by \citet{dotter11} and by \citet{bellini13}. 

The scenario emerging from the above empirical evidence is that 
metal--rich Galactic globulars appear to be, within the errors, 
coeval with metal--poor globulars \citep{monelli13,diceccoage}. 
This means that the quoted 
globulars do not show a clear evidence of an age metallicity 
relation. This is a preliminary conclusion, since we still lack 
homogeneous age estimates of a sizeable sample metal--rich 
clusters with a precision better than 1~Gyr. The fact that 
a significant fraction of metal--rich clusters are located inside 
or projected onto the Bulge is calling for new precise and 
deep NIR photometry. Together with the obvious bonus concerning 
the reduced impact of uncertainties in the cluster reddenings 
and the possible occurrence of differential reddening, there 
is also the advantage to fully exploit the precision of the 
new diagnostic ($\Delta^\mathrm{Knee}_\mathrm{TO}$).  

Finally, we would like to stress that the above findings 
appear also very promising concerning the solid evidence 
of a dichotomic age distribution of Galactic globulars. 
This evidence was brought forward by \citet{sala02}
and subsequently soundly confirmed by \citet{marin09}
an more recently, by \citet{vandenberg13} and 
\citet{leaman13}. The use of the $\Delta^\mathrm{Knee}_\mathrm{TO}$ parameter 
appears even more promising in this context appears even more 
promising. The reason is twofold. 
a) empirical evidence indicate that the morphology of the 
MSK is well defined when moving from the metal-rich 
\citep[NGC~6528][in preparation]{sara07,lagioia} to the 
metal--intermediate (Bono et al. 2010) and to the metal-poor 
regime \citep[M15,][]{monelli15}.     
b) Relative ages based on the $\Delta^\mathrm{Knee}_\mathrm{TO}$ are less prone 
to systematics, since the MSK is independent of cluster age and 
mildly affected by changes in helium content. The vertical 
method that is based on the difference between the HB luminosity 
level and the MSTO is more prone to the quoted possible systematics. 
Indeed, the HB luminosity level is affected by both cluster age and
helium content.     
     
The suggested experiment appears to be luckily supported by the unique 
opportunity to use in the near future adaptive optics systems at the 
8~m class telescope with superb image quality and spatial resolution 
\citep{schreiber,fiore14}.


\acknowledgments
This investigation was partially supported by PRIN-MIUR (2010LY5N2T) "Chemical and dynamical evolution of the Milky Way and Local Group galaxies" (P.I.: F. Matteucci).

It is a pleasure to acknowledge the anonymous referee for his/her very pertinent
suggestions on the early version of our paper.

One of us, A.D.C. thanks the ASI Science Data Center, and in particular Dr. Lucio Angelo Antonelli for supporting this investigation; while G.B. thanks the Carnegie Observatory visitor programme for support as science visitor. 

This research used the facilities of the Canadian Astronomy Data Centre operated by the National Research Council of Canada with the support of the Canadian Space Agency.

This publication makes use of data products from the Two Micron All Sky Survey, which is a joint project of the University of Massachusetts and the Infrared Processing and Analysis Center/California Institute of Technology, funded by the National
Aeronautics and Space Administration and the National Science Foundation.
 

\bibliographystyle{apj}

\LongTables
\begin{deluxetable}{cccccc}
\tablewidth{0pt}
\tabletypesize{\footnotesize}
\tablecaption{Ridgelines of M71 in the $r',u'-g'$ and $r',g'-r'$ CMDs\tablenotemark{a}.}
\tablehead{
$r'$ & $\sigma_{r'}$ & $u'-g'$ & $\sigma_{u'-g'}$ & $g'-r'$ & $\sigma_{g'-r'}$ \\
mag &~mag &~mag & mag &~mag & mag }
\startdata
11.49  & 0.01 &  3.94 & 0.04  & 1.58 & 0.12   \\  
11.69  & 0.01  &  3.70 & 0.04  & 1.47  & 0.04   \\  
11.89  & 0.01  &  3.48 & 0.07  & 1.38  & 0.09   \\  
12.09  & 0.01  &  3.30 & 0.06  & 1.32  & 0.10   \\  
12.21  & 0.01  &  3.21 & 0.06  & 1.28  & 0.06   \\  
12.33  & 0.01  &  3.13 & 0.04  & 1.25  & 0.04   \\  
12.50  & 0.01  &  3.02 & 0.03  & 1.21  & 0.03   \\  
12.66  & 0.01  &  2.92 & 0.08  & 1.18  & 0.05   \\  
12.82  & 0.01  &  2.83 & 0.06  & 1.15  & 0.04   \\  
13.02  & 0.01  &  2.72 & 0.03  & 1.11  & 0.03   \\  
13.22  & 0.01  &  2.62 & 0.04  & 1.08  & 0.03   \\  
13.46  & 0.01  &  2.52 & 0.03  & 1.05  & 0.04   \\  
13.70  & 0.01  &  2.42 & 0.03  & 1.01  & 0.03   \\  
13.94  & 0.01  &  2.32 & 0.03  &    0.98   & 0.02  \\  
14.18  & 0.01  &  2.24 & 0.02  &    0.96   & 0.02  \\  
14.42  & 0.01  &  2.16 & 0.02  &    0.93   & 0.02  \\  
14.67  & 0.01  &  2.09 & 0.02  &    0.91   & 0.02  \\  
14.91  & 0.01  &  2.03 & 0.02  &    0.89   & 0.02  \\  
15.15  & 0.01  &  1.98 & 0.02  &    0.87   & 0.02  \\  
15.39  & 0.01  &  1.94 & 0.02  &    0.86   & 0.02  \\  
15.63  & 0.01  &  1.89 & 0.02  &    0.84   & 0.02  \\  
15.91  & 0.01  &  1.85 & 0.02  &    0.83   & 0.02  \\ 
16.05  & 0.01  &  1.83 & 0.02  &    0.82   & 0.02  \\  
16.19  & 0.01  &  1.81 & 0.02  &    0.82   & 0.02  \\  
16.33  & 0.01  &  1.80 & 0.01  &    0.81   & 0.02  \\  
16.51  & 0.01  &  1.78 & 0.01  &    0.81   & 0.01  \\  
16.63  & 0.01  &  1.76 & 0.01  &    0.80   & 0.01  \\  
16.71  & 0.01  &  1.75 & 0.01  &    0.80   & 0.01  \\  
16.77  & 0.01  &  1.74 & 0.01  &    0.80   & 0.01  \\  
16.83  & 0.01  &  1.74 & 0.01  &    0.80   & 0.01  \\  
16.87  & 0.01  &  1.73 & 0.01  &    0.79   & 0.01  \\  
16.91  & 0.01  &  1.72 & 0.01  &    0.79   & 0.01  \\  
16.95  & 0.01  &  1.72 & 0.01  &    0.79   & 0.01  \\  
16.99  & 0.01  &  1.71 & 0.01  &    0.78   & 0.01  \\  
17.03  & 0.01  &  1.70 & 0.01  &    0.78   & 0.01  \\  
17.07  & 0.01  &  1.69 & 0.01  &    0.77   & 0.01  \\  
17.11  & 0.01  &  1.68 & 0.01  &    0.76   & 0.01  \\  
17.15  & 0.01  &  1.66 & 0.01  &    0.75   & 0.01  \\  
17.17  & 0.01  &  1.64 & 0.01  &    0.74   & 0.01  \\  
17.19  & 0.01  &  1.61 & 0.01  &    0.73   & 0.01  \\  
17.21  & 0.01  &  1.58 & 0.01  &    0.72   & 0.01  \\  
17.23  & 0.01  &  1.53 & 0.01  &    0.71   & 0.01  \\  
17.25  & 0.01  &  1.48 & 0.01  &    0.70   & 0.01  \\  
17.27  & 0.01  &  1.44 & 0.01  &    0.69   & 0.01  \\  
17.29  & 0.01  &  1.41 & 0.01  &    0.68   & 0.01  \\  
17.31  & 0.01  &  1.38 & 0.01  &    0.67   & 0.01  \\  
17.33  & 0.01  &  1.35 & 0.01  &    0.65   & 0.01  \\  
17.35  & 0.01  &  1.33 & 0.01  &    0.65   & 0.01  \\  
17.37  & 0.01  &  1.32 & 0.01  &    0.64   & 0.01  \\  
17.39  & 0.01  &  1.31 & 0.01  &    0.63   & 0.01  \\  
17.43  & 0.01  &  1.29 & 0.01  &    0.62   & 0.01  \\  
17.47  & 0.01  &  1.27 & 0.01  &    0.62   & 0.01  \\  
17.51  & 0.01  &  1.26 & 0.01  &    0.61   & 0.01  \\  
17.55  & 0.01  &  1.25 & 0.01  &    0.61   & 0.01  \\  
17.57  & 0.01  &  1.24 & 0.01  &    0.60   & 0.01  \\  
17.59  & 0.01  &  1.24 & 0.01  &    0.60   & 0.01  \\  
17.61  & 0.01  &  1.24 & 0.01  &    0.60   & 0.01  \\  
17.63  & 0.01  &  1.24 & 0.01  &    0.60   & 0.01  \\  
17.65  & 0.01  &  1.23 & 0.01 &    0.60    & 0.01 \\  
17.67  & 0.01  &  1.23 & 0.01  &    0.60   & 0.01  \\  
17.71  & 0.01  &  1.23 & 0.01	&    0.60  & 0.01   \\  
17.75  & 0.01 &  1.23  & 0.01 &    0.60    & 0.01 \\  
17.79  & 0.01 &  1.23  & 0.01  &    0.60   & 0.01  \\  
17.83  & 0.01 &  1.23  & 0.01  &    0.60   & 0.01  \\  
17.89  & 0.01  &  1.23 & 0.01  &    0.60   & 0.01\\  
17.95  & 0.01  &  1.24 & 0.01  &    0.60  & 0.01\\    
18.07  & 0.01 &  1.25  & 0.01 &    0.60    & 0.01 \\  
18.23  & 0.01 &  1.27  & 0.01  &    0.61   & 0.01  \\  
18.39  & 0.01  &  1.30 & 0.01  &    0.62   & 0.01  \\  
18.55  & 0.01  &  1.33 & 0.01  &    0.64   & 0.01  \\  
18.71  & 0.01  &  1.38 & 0.01  &    0.66   & 0.01  \\  
18.87  & 0.01  &  1.43 & 0.01  &    0.68   & 0.01  \\  
19.03  & 0.01 &  1.49  & 0.01 &    0.71    & 0.01 \\  
19.23  & 0.01  &  1.57 & 0.01  &    0.74   & 0.01  \\  
19.44  & 0.01  &  1.67 & 0.01  &    0.78   & 0.01  \\  
19.64  & 0.01  &  1.78 & 0.01  &    0.83   & 0.01  \\  
19.84  & 0.01  &  1.90 & 0.01  &    0.88   & 0.01  \\  
20.04  & 0.01  &  2.03 & 0.01  &    0.93   & 0.01  \\  
20.20  & 0.01  &  2.14 & 0.01  &    0.97   & 0.01  \\  
20.40  & 0.01 &  2.29  & 0.01 &     1.04   & 0.01  \\  
20.60  & 0.01  &  2.43 & 0.01  &     1.11  & 0.01   \\  
20.88  & 0.01  &  2.59 & 0.01  &     1.20  & 0.01   \\  
21.12  & 0.01  &  2.73 & 0.01  &     1.28  & 0.01   \\  
21.36  & 0.01  &  2.87 & 0.01 &     1.35   & 0.01  \\  
21.56  & 0.02  &  2.97 & 0.01  &     1.41  & 0.01   \\  
21.76  & 0.02  &  3.07 & 0.02  &     1.46  & 0.02   \\  
21.84  & 0.02  &  3.11 & 0.02  &     1.48  & 0.02   \\
\enddata
\tablenotetext{a}{The sampling in magnitude and in color along the $r'$,$u'-g'$ 
and the $r'$,$g'-r'$ ridgelines is not uniform. The sampling increases in the 
regions in which there are relevant changes in the slope (main sequence turn off, 
base of the red giant).  The $\sigma$ marks the uncertainty either in magnitude or in color (summed in 
quadrature).}
\label{tab1}
\end{deluxetable}


\begin{deluxetable}{ll lcl}
\tablewidth{0pt}
\tabletypesize{\footnotesize}
\tablecaption{Absolute ages, distance moduli, reddening estimates and iron abundances for M71.}
\tablehead{
DM$_V$\tablenotemark{a} & $E(\bmv)$\tablenotemark{b} & Age\tablenotemark{c}& \feh\tablenotemark{d} & Notes\tablenotemark{e} \\
                   mag  &                   mag   &                 Gyr &                         &                        
}
\startdata
13.70                   & 0.28          & 14$\pm 2$; 16$\pm 2$ & -0.78 & H92 \\                       
14.09$\pm$0.15          & 0.28          & 8$\pm$1              & -0.70 & R92 \\
13.60$\pm$0.10          & 0.27$\pm$0.05 & 18                   & -1.02 & G00 \\   
13.71$\pm$0.04$\pm$0.1  & 0.28          & 12                   & -0.70 & G02 \\ 
13.78                   & 0.20          & 11                   & -0.80 & B10 \\
13.69                   & 0.24          & $11.00\pm0.38$       & -0.82 & V13 \\
13.84                   & 0.25          & 12$\pm$2             & -0.78 & MSTO \\
\ldots                  & \ldots        & 12$\pm$1             & -0.78 & $\Delta^\mathrm{Knee}_\mathrm{TO}$ 
\enddata
\tablenotetext{a}{Apparent distance modulus and its error when estimated by the quoted authors.}
\tablenotetext{b}{Adopted cluster reddening.}
\tablenotetext{c}{Absolute cluster age.}
\tablenotetext{d}{Adopted iron abundance.}  
\tablenotetext{e}{Notes: 
H92, \citet{hodder92}; R92, \citet{reid98}; G00, \citet{geffert2000}; 
G02, \citet{grund02};  B10, \citet{brasseur10}; V13, \cite{vandenberg13}.
MSTO: age estimate based on the main sequence turn off; 
$\Delta^\mathrm{Knee}_\mathrm{TO}$: age estimate based on the difference in magnitude between MSTO and knee.}
\label{tab2}
\end{deluxetable}


\begin{deluxetable}{ll}
\tablewidth{0pt}
\tabletypesize{\footnotesize}
\tablecaption{Effective temperatures and gravities of the selected models to study the effect of CIA.}
\tablehead{
$T_\mathrm{eff}$ & $\log g$ \\
$[\mathrm{K}]$ & $[\mathrm{cgs}]$}
\startdata
  3777  & 4.98 \\
  3904  & 4.91\\
  4081  & 4.82
\enddata
\label{tab3}
\end{deluxetable}


\begin{deluxetable}{lc}
\tabletypesize{\footnotesize}
\tablewidth{0pt}
\tablecaption{Collisional Induced Absorption opacities included in the 
PHOENIX BT-Settl atmosphere models (allard et al. 2011).}
\tablehead{
CIA & Ref.}
\startdata
H$_2-$H$_2$  & \tablenotemark{1, 2} \\ 
H$_2-$He   & \tablenotemark{3, 4} \\ 
H$_2-$H   & \tablenotemark{5} \\ 
He$-$H     & \tablenotemark{6} \\ 
H$_2-$CH$_4$ & \tablenotemark{7, 8}\\ 
H$_2-$N$_2$  & \tablenotemark{9, 10}\\ 
N$_2-$CH$_4$ & \tablenotemark{11, 12}\\ 
N$_2-$N$_2$  & \tablenotemark{13,14}\\ 
CH$_4-$CH$_4$ & \tablenotemark{15}\\ 
CO$_2-$CO$_2$ & \tablenotemark{16, 17}\\  
H$_2-$Ar   & \tablenotemark{18, 19}\\ 
CH$_4-$Ar  & \tablenotemark{20, 21}\\  
\enddata
\tablenotetext{1}{\citet{bory02};
$^{2}$ \citet{abel11};  
$^{3}$ \citet{bory89};
$^{4}$ \citet{bory97}; 
$^{5}$ \citet{gusta03};
$^{6}$ \citet{gusta01};
$^{7}$ \citet{bory86H2CH4};
$^{8}$ \citet{boryfromdore86};
$^{9}$ \citet{bory86H2N2};
$^{10}$ \citet{dore86};
$^{11}$ \citet{borytang};
$^{12}$ \citet{Birnbaum93};
$^{13}$ \citet{bory86_3};
$^{14}$ \citet{bory87N2N2};
$^{15}$ \citet{bory87CH4CH4};
$^{16}$ \citet{gru97};
$^{17}$ \citet{gru98};
$^{18}$ \citet{mey86};
$^{19}$ \citet{borymora94};
$^{20}$ \citet{dore90};
$^{21}$ \citet{borymora93}. 
}
\label{tab4}
\end{deluxetable}

%



\end{document}